\documentclass[conference]{IEEEtran}
\IEEEoverridecommandlockouts
\usepackage{cite}
\usepackage{amsmath,amssymb,amsfonts}
\usepackage{algorithmic}
\usepackage{graphicx}
\usepackage{textcomp}
\usepackage{xcolor}
\usepackage{calligra}
\usepackage{algorithm}
\usepackage{algorithmic}

\usepackage{caption}
\usepackage{float} 
\usepackage{subfigure}
\usepackage{makecell}

\def\BibTeX{{\rm B\kern-.05em{\sc i\kern-.025em b}\kern-.08em
    T\kern-.1667em\lower.7ex\hbox{E}\kern-.125emX}}

\begin{document}

\title{Can We Improve Channel Reciprocity via Loop-back Compensation for RIS-assisted Physical Layer Key Generation}

\author{
\thanks{*Guoshun Nan is the corresponding author}
\IEEEauthorblockN{
Ningya Xu\IEEEauthorrefmark{2},
Guoshun Nan\IEEEauthorrefmark{1}\IEEEauthorrefmark{2},
Xiaofeng Tao\IEEEauthorrefmark{2}, Na Li\IEEEauthorrefmark{2}, Pengxuan Mao\IEEEauthorrefmark{3}, Tianyuan Yang\IEEEauthorrefmark{3}
\IEEEauthorblockA{\IEEEauthorrefmark{2}Beijing University of Posts and Telecommunications
}
\IEEEauthorblockA{\IEEEauthorrefmark{3}Terminus Technologies Co., Ltd
}
\IEEEauthorblockA{xuningya2017@bupt.edu.cn, nanguo2021@bupt.edu.cn, taoxf@bupt.edu.cn, nali\_bupt@163.com, \\mao.pengxuan@tslsmart.com,  yang.tianyuan@tslsmart.com}
}
}

\maketitle

\begin{abstract}
Reconfigurable intelligent surface (RIS) facilitates the extraction of unpredictable channel features for physical layer key generation (PKG), securing communications among legitimate users with symmetric keys. Previous works have demonstrated that channel reciprocity plays a crucial role in generating symmetric keys in PKG systems, whereas, in reality, reciprocity is greatly affected by hardware interference and RIS-based jamming attacks. This motivates us to propose LoCKey, a novel approach that aims to improve channel reciprocity by mitigating interferences and attacks with a loop-back compensation scheme, thus maximizing the secrecy performance of the PKG system. Specifically, our proposed LoCKey is capable of effectively compensating for the CSI non-reciprocity by the combination of transmit-back signal value and error minimization module. Firstly, we introduce the entire flowchart of LoCKey and provide an in-depth discussion of each step. Following that, we delve into a theoretical analysis of the performance optimizations when our LoCKey is applied for CSI reciprocity enhancement. Finally, we conduct experiments to verify the effectiveness of the proposed LoCKey in improving channel reciprocity under various interferences for RIS-assisted wireless communications. The results demonstrate a significant improvement in both the rate of key generation assisted by the RIS and the consistency of the generated keys, showing great potential for 
the practical deployment of our LoCKey in future wireless systems.
\end{abstract}


\section{Introduction}
Due to the openness of wireless channels and the continuous development of attack methods, severe security problems faced by wireless communication have become increasingly prominent\cite{jxs}. According to research by technology company McAfee, millions of malicious Wi-Fi attacks occur every day, potentially leading to serious privacy breaches. Traditional symmetric encryption requires the pre-distribution of keys between each communicator, which becomes extremely complex and impractical in Internet of Things (IoT) scenarios with a huge number of terminal equipment\cite{lgy1}. The goal of physical layer key generation (PKG) is to explore the endogenous security mechanism of wireless communication by using the channel inherent characteristics such as time-variability\cite{hkz} and channel reciprocity\cite{liuh}, to provide lightweight encryption. The natural properties of wireless channels make PKG have the unique advantages of low cost and high speed in solving distribution and renewal of secret keys.

However, the non-obvious Doppler effect in the quasi-static scenario of the IoT leads to slow channel change\cite{lgy2}, significantly degrading the key generation rate (KGR). In recent years, Reconfigurable Intelligent Surface (RIS) has been a new technology that can adjust electromagnetic waves, to customize the wireless propagation environment \cite{wq}. Thanks to the characteristics of passive elements and low hardware cost, RIS-assisted wireless communication systems had been widely studied. A closed expression for the upper bound of KGR was derived in \cite{lt}, which showed the relation of KGR to the number of RIS elements, correlation coefficient, pilot length, etc. Lu et.al\cite{lx} discussed how to maximize KGR by using channel state information (CSI) to dynamically adjust the open position of RIS components. It was verified in \cite{jl} that KGR with RIS assistance can be increased by 197.5\% compared with the relay-assisted scheme. RIS can also facilitate PKG in mmwave system \cite{sy} or vehicle-to-vehicle (V2V) communications\cite{ayaz}.

The above studies consider how to minimize the effect of passive eavesdropping, while the possibility of RIS being attacked by malicious external attackers remains under-explored. Normal jamming attacks in PKG expressed by sending interference signals to the effective frequency of the channel detection\cite{zafer}. Specifically, RIS-jamming is represented by Mallory alienating the RIS reflection matrix of upstream and downstream channels\cite{lgy4}. If RIS applies a random configuration at an update rate higher than the channel sampling rate, the difference in CSI reciprocity results in unsuccessful key negotiation. Lyu et.al\cite{lyu} proposed the use of RIS as a green jammer to attack communications without using any internal energy, Hu et.al\cite{lh} proposed an attack strategy based on RIS and a multipath detection separation scheme based on a broadband system, but the randomness enhancement brought by RIS cannot be used in the key generation process. Unlike traditional jamming attacks, RIS jamming attacks do not emit but reflect interference signals, so the location of "Eve" is difficult to reveal, and traditional countermeasures based on position detection\cite{eberz} will be ineffective. What's more, the reason for the non-reciprocity of CSI in Time Division Duplexing (TDD) systems may also be caused by hardware fingerprint interference\cite{lp} in practical systems. Current research mainly uses pre-coding techniques to compensate for non-reciprocity\cite{jz}, which requires a complex learning process and is not suitable for key generation scenarios that require the ability of real time channel information obtaining.

Therefore, our paper considers the perspective of channel reciprocity compensation. We put forward LoCKey, a novel scheme that can not only remove non-reciprocity caused by interference but also apply the randomness brought by RIS to improve the key generation rate. The main contributions of this paper can be summarized as follows:\\
$\bullet$ Our proposed LoCKey, as a novel channel compensation method that aims to improve the reciprocity of the RIS-assisted physical layer key generation, can facilitate the rapid changes in the phase of RIS to manipulate the up-downward channels. \\
$\bullet$ We introduce the overall process of the proposed LoCKey. We also present a theoretical analysis to show how LoCKey improves the channel reciprocity, aiming to effectively mitigate hardware interference and RIS-based jamming attacks.\\
$\bullet$ Experiments show the superiority of LoCKey compared to existing loop-back schemes and non-loop-back ones.

\section{System Model}

\subsection{General PKG Procedure}

\paragraph{Channel Measurement}
Alice and Bob send pilot frequency to the peer end, and generate a random key source by probing some characteristics of the channel, e.g. received signal strength(RSS), CSI, channel phase response, and channel multipath delay. In our research, we adopt CSI as the key source.

\paragraph{Quantization}

The communication parties convert the measured value of the channel into a bit sequence of 0 and 1. 

\paragraph{Information Reconciliation}

Discard or correct the differences between key bit streams caused by device differences\cite{zs} or additive noise, reducing the inconsistency rate. 

\paragraph{Privacy Amplification}

The sequence after information reconciliation is mapped to produce a shorter random key, exponentially reducing the leaked information to the eavesdropper\cite{rd}. The cryptographic Hash algorithms (Digest algorithms) are usually used for privacy amplification.

As our paper researches the performance of our channel compensation method brought to the key generation rate and bit difference rate, we focus mostly on the \textit{Channel Measurement} step and \textit{Quantization} step.

\subsection{Our RIS-assisted PKG model}

\begin{figure}[htbp]
\centerline{\includegraphics[width=0.7\linewidth]{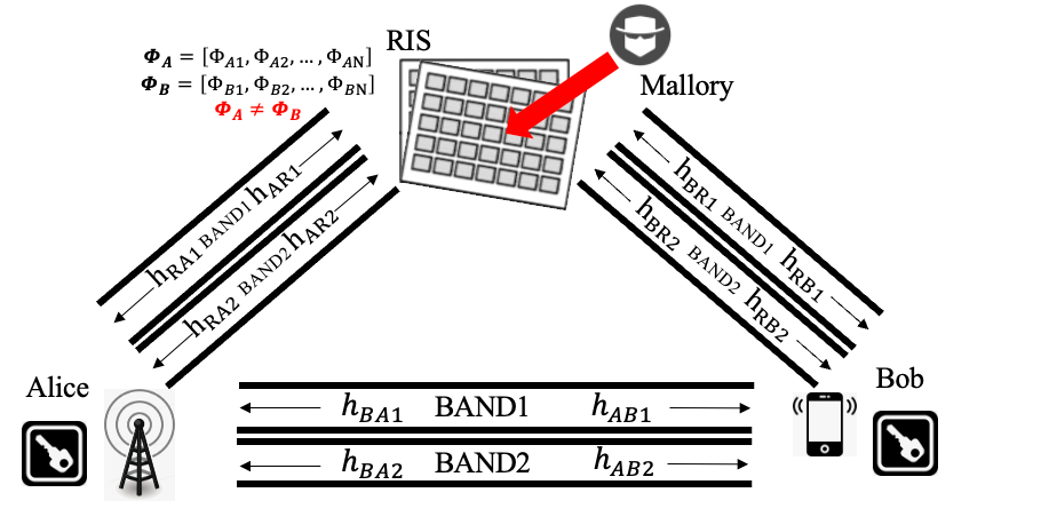}}
\caption{RIS-assisted PKG system model with attacked RIS.}
\label{2-1}
\end{figure}

We consider a general RIS-assisted PKG system shown in Fig.~\ref{2-1}.  We suppose two different frequency bands (BAND 1 and BAND 2), and the transceiver can selectively turn on or off several subcarriers for each transmission to transmit through different bands, which is suitable for TDD systems with multi-carrier modulation. 

As Fig.~\ref{2-1} shows, single-antenna-equipped legitimate transceivers Alice and Bob take turns to transmit OFDM symbols $ \textbf{s}=diag(s_1, s_2, ... s_L) $ within the coherence time $\tau$ during channel probing stage,  where $L$ is the number of subcarriers. A RIS consisting of $N$ reflecting units is used for channel randomness improvement, assuming that all units are independent, and the status value $\omega$ can be set to ``on" $\omega=1$ or ``off" $\omega=0$ by the controller. We change the phase shift matrix $\varphi=[\omega_1\phi_1,\omega_2\phi_2,\omega_3\phi_3,..., \omega_N\phi_N]^T,\, \omega_i\in\{0, 1\}$ in real time, where $\phi_i\in(0,2\pi)$ denotes the random phase shift corresponding to each RIS unit. Unavoidable hardware deviation interference -which we write as $\mathrm{F_d}$, where $\mathrm{d}$ is the transmission direction- is a factor that causes CSI non-reciprocity.

The received signal strength (RSS) on both sides $y_{A_i}$ and $y_{B_i}$ can be written as follows, where $i\in\{1,2\}$ denote the BAND used for transmission, $n$ could be seen as additive white Gaussian noise (AWGN) of each transmission.

\begin{equation}
\begin{aligned}
y_{A_i}= \mathrm{F_{BA}}(h_{BA_i}+ \sum_{i=1}^N h_{BR_i} \varphi_{A_i}  h_{RA_i}) \textbf{s} +n\\
y_{B_i}= \mathrm{F_{AB}}(h_{AB_i}+ \sum_{i=1}^N h_{AR_i} \varphi_{B_i}  h_{RB_i}) \textbf{s} +n
\label{eq1}
\end{aligned}
\end{equation}

By using the least square (LS) channel estimation method to RSS values, the CSI values at Alice $H_{A_i}$ and Bob $H_{B_i}$ could be expressed as follows, mentioned that the RIS-cascaded channel can be written as the multiplication of two sub-channels through RIS, i.e. $h_{ARB_i}=h_{AR_i} \cdot h_{RB_i}$, $h_{BRA_i}=h_{BR_i} \cdot h_{RA_i}$. In the calculation process, we write $\varPhi_A = \sum_{i=1}^N \varphi_A$, $\varPhi_B = \sum_{i=1}^N \varphi_B$.

\begin{equation}
\begin{aligned}
{H}{_{A_i}}= \mathrm{F_{BA}}(h_{BA_i}+ h_{BRA_i} \varPhi_A)+n\\
{H}{_{B_i}}= \mathrm{F_{AB}}(h_{AB_i}+ h_{ARB_i} \varPhi_B)+n
\label{eq2}
\end{aligned}
\end{equation}

RIS active jamming attack aims at disrupting the key establishment process between Alice and Bob by adjusting the RIS reflection matrices. A malicious external attacker Mallory launches active jamming by randomly changing the RIS reflection matrices in the upward and downward RIS-induced link, causing $ \varPhi_A \ne \varPhi_B $. As long as the update rate of RIS configurations is higher than the channel sampling rate, the observed CSI of RIS-induced links appear to be different, therefore destroying the total channel reciprocity. Note that the RIS matrix after being attacked will change randomly over time, while the hardware deviation $\mathrm{F_d}$ is a constant value.

\section{Our LoCKey Scheme}

\subsection{Scheme Flowchart}

Under our attacked-RIS PKG model, we propose the LoCKey Transmission Scheme shown in Fig.~\ref{2-2}, in order to strengthen the reciprocity of the CSI values:

\begin{figure}[htbp]
\centerline{\includegraphics[width=0.8\linewidth]{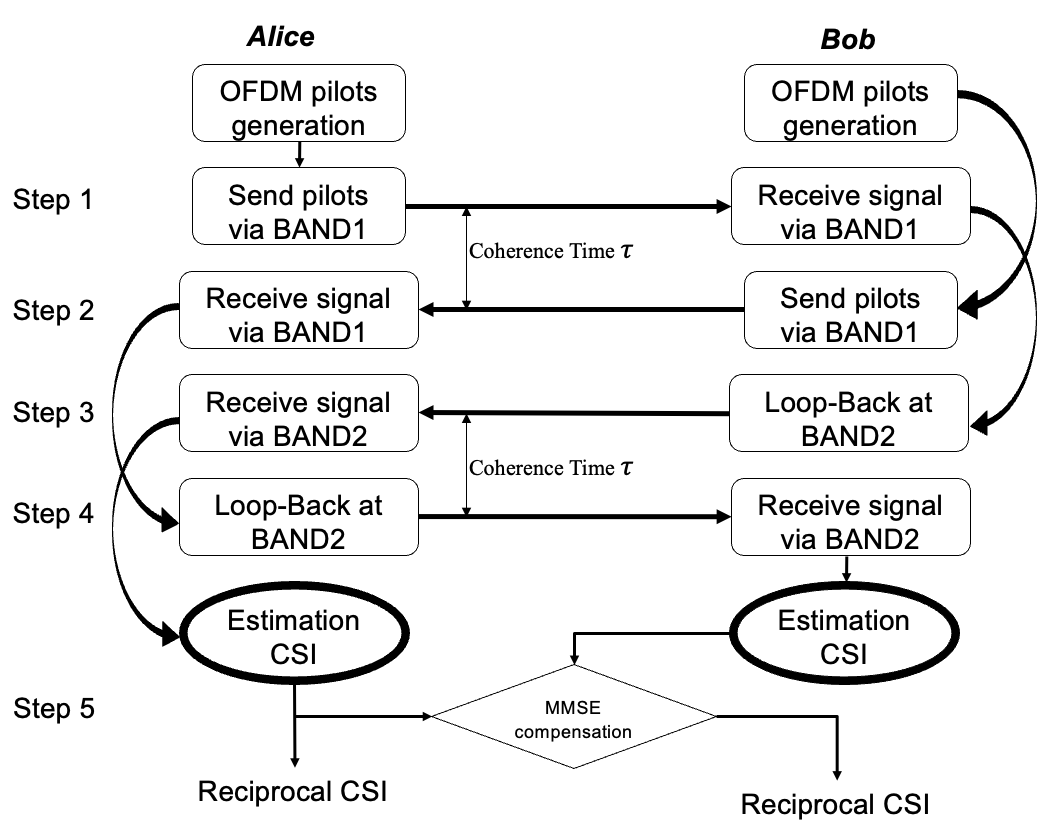}}
\caption{Flowchart of the proposed LoCKey. }
\label{2-2}
\end{figure}

Step 1: Alice first sends a common OFDM pilot signal to Bob through BAND 1. 

Step 2: Bob sends a pilot signal to Alice through BAND 1. The period between Step 1 and Step 2 is a coherence time $\tau$.

Step 3: After the first round of channel measurements, Bob sends the signal received in Step 1 out through frequency BAND 2 at time $t+\tau$, and Alice can receive the loop signal.

Step 4: Alice sends the signal received in Step 2 out through BAND 2, and Bob receives the loop signal. The period between Step 3 and Step 4 is a coherence time $\tau$.

Step 5: In this module, we measure the error between the actual measured value at Bob in Step 4 and the predicted value at Alice in Step 3, then minimize the MSE of the prediction by reasonable selection of the predicted scalar value.

Step 1 to 2 is the classical TDD process, and Step 3 to 4 is the added loop-back process for the PKG purpose, and Step 5 is our channel reciprocity enhanced module. Therefore, the whole transmission scheme is called as LoCKey scheme.

\subsection{Scheme Description}

Channel changes between measurements in TDD systems are difficult to model accurately, so we assume that the change of CSI is continuously integrable for simplicity. Considering the selection of frequency bands $i,j\in\{1,2\}$, we define the sub-channels CSI performance under different frequency bands as follows:
\begin{equation}
h^t_{u_i}  \ne h^t_{u_j}, i \ne j 
\label{eq3}
\end{equation}

At Step 1 of our LoCKey scheme, the received CSI by Bob from Alice is expressed by:
\begin{equation}
\begin{aligned}
{H}{^t_{B_1}}= \mathrm{F_{AB}}(h^t_{AB_1}+ h^t_{ARB_1} \varPhi^t_B)+n
\end{aligned}
\label{e5}
\end{equation}

The legitimate parties send the pilot signal only once within each coherence time $\{t,t+\tau \}$, then they receive the pilot signal and estimate the channel. It is assumed that the channels in coherence time remain constant and the legitimate channels satisfy perfect reciprocity:
\begin{equation}
\begin{aligned}
h_{AB_i} \approx h_{BA_i} ,h_{ARB_i} \approx h_{BRA_i}
\label{eq5}
\end{aligned}
\end{equation}

Therefore in Step 2, the received CSI by Alice from Bob after a channel detection turn can be presented as follows by using the channel reciprocity equation in (5):
\begin{equation}
\begin{aligned}
{H}{^{t}_{A_1}} &= \mathrm{F_{BA}}(h^{t}_{BA_1}+ h^{t}_{BRA_1} \varPhi^{t}_A) +n\\
& \approx \mathrm{F_{BA}}(h^t_{AB_1}+ h^t_{ARB_1} \varPhi^{t}_A) +n
\end{aligned}
\label{e6}
\end{equation}

After the first round of channel measurements, by sending the loop-back CSI in Step 1 to Step 3, we get a mutual CSI of the two frequency bands at Alice:
\begin{equation}
\begin{aligned}
{H}^{t+\tau}{_{A}} &= {H}{^{t}_{B_1}}{H}{^{t+\tau}_{A_2}} \\
&=\mathrm{F_{BA}}(h^{t+\tau}_{BA_2}+ h^{t+\tau}_{BRA_2} \varPhi^{t+\tau}_A)[\mathrm{F_{AB}}
(h^{t}_{AB_1}+ h^{t}_{ARB_1}\\ & \varPhi^t_B) +n]+n\\
&=\mathrm{F_{BA}}\mathrm{F_{AB}}(h^{t+\tau}_{BA_2}+ h^{t+\tau}_{BRA_2} \varPhi^{t+\tau}_A)(h^t_{AB_1} + h^t_{ARB_1} \\
&\varPhi^t_B) + \mathrm{F_{BA}}(h^{t+\tau}_{BA_2}+ h^{t+\tau}_{BRA_2} \varPhi^{t+\tau}_A)n+n
\label{e7}
\end{aligned}
\end{equation}

And by sending the loop-back CSI in Step 2 to Step 4, we get a mutual CSI at Bob, also using the channel reciprocity equation in \eqref{eq5}:
\begin{equation}
\begin{aligned}
{H}^{t+\tau}{_{B}}&={H}{^{t}_{A_1}}{H}{^{t+\tau}_{B_2}}\\
&= \mathrm{F_{AB}}\mathrm{F_{BA}}(h^{t+\tau}_{AB_2}+ h^{t+\tau}_{ARB_2} \varPhi^{t+\tau}_B)(h^{t}_{BA_1}+  h^{t}_{BRA_1} \\
& \varPhi^{t}_A)+ \mathrm{F_{AB}}(h^{t+\tau}_{AB_2}+ h^{t+\tau}_{ARB_2} \varPhi^{t+\tau}_B)n+n\\
&\approx \mathrm{F_{AB}}\mathrm{F_{BA}}(h^{t+\tau }_{BA_2}+ h^{t+\tau}_{BRA_2} \varPhi^{t+\tau}_B)(h^{t}_{AB_1}+  h^{t}_{ARB_1} \\
& \varPhi^{t}_A)+ \mathrm{F_{AB}}(h^{t+\tau}_{BA_2}+ h^{t+\tau}_{BRA_2} \varPhi^{t+\tau}_B)n+n\\
\label{e8}
\end{aligned}
\end{equation}

We adopt ${H}^{t+\tau}_{B}$ and ${H}^{t+\tau}_{A}$ as key source for the future secret key generation steps. From the above two equations, we can see that the non-reciprocity brought by hardware deviation is eliminated, by sharing common randomness of cross multiplication $\mathrm{F_{AB}}\mathrm{F_{BA}}$. 

But as we see, the influence of RIS-jamming still exists because the manipulated RIS matrix $\varPhi_A$ and $\varPhi_B$ are unpredictable and independent due to the malicious attacking of Mallory. 

That's why we introduce Step 5, an MMSE compensation module. In order to compensate for the imperfect channel reciprocity, the transmitter needs to predict the wireless channel that would be experienced by the receiver, particularly the channel estimation values that would be obtained by the receiver. The prediction follows the MMSE prediction methodology, where $\Gamma (k)$ is a predicted scalar value for the channel prediction of the $k^{th}$ subcarrier\cite{hl}.

\begin{equation}
\begin{aligned}
     \widetilde {H}^{t+\tau}_B(k)= \Gamma(k) \widetilde{H}^{t+\tau}_A(k), k=0,1,...,L-1
\end{aligned}
\label{e9}
\end{equation}

The error between the actual measured value at Bob and the predicted value at Alice in Step 3 is expressed as $\varepsilon(k)$ by calculating the error between the actual measured value at Bob when the $k^{th}$ subcarrier is sent after Step 4. We consider that $\widetilde{H}^{t+\tau}_A(k)=H^{t+\tau}_A(k)$ here for the ease of calculation.

\begin{equation}
\begin{aligned}
    \varepsilon(k)&=\widetilde {H}^{t+\tau}_B(k)-H^{t+\tau}_B(k)\\
    &=\Gamma(k)H^{t+\tau}_A(k)-H^{t+\tau}_B(k)
\end{aligned}
\label{e10}
\end{equation}

The mean squared error (MSE) of the channel prediction can then be derived as $E[\varepsilon(k)^2]$, then the MSE of the prediction is minimized by a reasonable selection of the predicted scalar value $\Gamma(k)$ representing the $k^{th}$ subcarrier of the channel prediction, i.e. $min E[\varepsilon(k)^2]$. 

After Step 5, the non-reciprocal channels caused by RIS-jamming can be adjusted for recovery.
\subsection{Performance Analysis}

In this section, we verify the above architecture by calculating the correlation coefficient between the channel estimation values. The correlation coefficient between two random variables X and Y can be calculated by the following formula:
\begin{equation}
\begin{aligned}
    \rho &=\frac{Cov(X,Y)}{\sigma X * \sigma Y}\\
    &=\frac{E[XY^*]-E[X]E[Y^*]}{\sqrt{E[|X|^2]}  \sqrt{E[|Y|^2]}}
\end{aligned}
\label{e11}
\end{equation}

\noindent where $E[~ ]$ denotes the expectation operator.

Note that all sub-channels follow the Gaussian distribution, i.e. $h \sim CN(0,\sigma^2_h)$, and the background noise is expressed by $n \sim CN(0,\sigma^2_n)$. The uncorrelated RIS matrix expressed by $\varPhi_A=\sum_{i=1}^N \omega_{A_i}\phi_{A_i}$ and $\varPhi_B=\sum_{i=1}^N \omega_{B_i}\phi_{B_i}$, where $\omega_{A_i}$ and $\omega_{B_i}$ are unmatched random values in the range of $(1,N)$, and $\phi$ is uniformly distributed between $(0,2\pi)$.

Therefore, after Step 1 and Step 2, the correlation coefficient between $H^t_{A_1}$ and $H^t_{B_1}$ can be expressed by:

\begin{equation}
\begin{aligned}
    \rho_1 &=\frac{E[H^t_{A_1} H^{t*}_{B_1}]-E[H^t_{A_1}]E[H^{t*}_{B_1}]}{\sqrt{E[|H^t_{A_1}|^2]} \sqrt{E[|H^{t}_{B_1}|^2]}}\\
    &=\frac{G_A G_B (ab\sigma^2_{ARB}+\sigma^2_{AB})}{\sqrt{G_A(a^2 \sigma^2_{ARB}+\sigma^2_{AB})+1}\sqrt{G_B(b^2\sigma^2_{ARB}+\sigma^2_{AB})+1}}
\end{aligned}
\label{e12}
\end{equation}

Whereas $G_A=E[\mathrm{F_{AB}} \mathrm{F^h_{AB}}]$, $G_B=E[\mathrm{F_{BA}} \mathrm{F^h_{BA}}]$, $a$ and $b$ are the statistical mean of RIS matrix in up-downward links. In the above equations, we assume that $\sigma^2_n=1$ for all sub-channels for better analysis of our target.

Taking the same approach, the correlation coefficient after Step 3 and Step 4 between the partly randomness-shared  $H^{t+\tau}_{A_1}$ and $H^{t+\tau}_{B_1}$ is:
\begin{equation}
\begin{aligned}
    \rho_2 &=\frac{E[H^{t+\tau}_{A} H^{t+\tau*}_{B}]-E[H^{t+\tau}_{A}]E[H^{t+\tau*}_{B}]}{\sqrt{E[|H^{t+\tau}_{A}|^2]}  \sqrt{E[|H^{t+\tau}_{B}|^2]}}\\
    &=\frac{G_A G_B (a^2 b^2(\sigma^4_{ARB})+(\sigma^4_{AB}))+1}{\sqrt{G_A(a^2\sigma^4_{ARB}+\sigma^4_{AB})+1}
    \sqrt{G_B (b^2\sigma^4_{ARB}+\sigma^4_{AB})+1}}
\end{aligned}
\label{e13}
\end{equation}

We apply an MMSE predictor in (9) to compensate for the decorrelation brought by manipulated RIS. Our goal is to minimize the mean squared error (MSE) of the channel prediction and the original channel estimation, therefore the MSE of channel prediction can then be derived as:

\begin{equation}
\begin{aligned}
   E[\varepsilon(k)^2]
    &=E[(\Gamma(k)H^{t+\tau}_A(k)-H^{t+\tau}_B(k))^2]\\
    &=\Gamma(k)^2[G_A(a^2\sigma^4_{ARB}+\sigma^4_{AB})+1]+[G_B (b^2\sigma^4_{ARB}+\\
    &\sigma^4_{AB})
    +1]-2\Gamma(k)[G_A G_B (a^2 b^2\sigma^4_{ARB}+
    \sigma^4_{AB})+1]
\end{aligned}
\label{e14}
\end{equation}

When the derivative of a function at a certain point is zero, this point may be a local minimum (or maximum) of the function. Therefore, it is necessary to check whether the derivative of the sum of squared error function with respect to $\Gamma(k)$ is zero to determine the optimal fitting parameters, which means that the derivative of $E[\varepsilon(k)^2]$with respect to $\Gamma(k)$ must be zero in order to minimize the sum of squared errors with respect to $\Gamma(k)$:
\begin{equation}
\begin{aligned}
   \frac{d}{d \Gamma(k)} E[\varepsilon(k)^2]=0
\end{aligned}
\label{e15}
\end{equation}

i.e.
\begin{equation}
\begin{aligned}
   & 2 \Gamma(k)\left[G_{A}\left(a^{2} \sigma_{A R B}^{4}+\sigma_{A B}^{4}\right)+1\right]\\
   & -2\left[G_{A} G_{B}\left(a^{2} b^{2} \sigma_{A R B}^{4}+\sigma_{A B}^{4}\right)+1\right]=0
\end{aligned}
\label{e16}
\end{equation}

Then the prediction scalar for the $k^{th}$ subcarrier can be derived as:
\begin{equation}
\begin{aligned}
   \rho_3= \Gamma(k)=\frac{G_A G_B (a^2 b^2\sigma^4_{ARB}+
    \sigma^4_{AB})+1}{G_A(a^2\sigma^4_{ARB}+\sigma^4_{AB})+1}
\end{aligned}
\label{e17}
\end{equation}

Finally, using $\widetilde {H}^{t+\tau}_B(k)$ and the reference CSI $\widetilde {H}^{t+\tau}_A(k)$ as key sources for subsequent key generation, the correlation coefficient after LoCKey is $\rho_3$. Since $\widetilde {H}^{t+\tau}_B(k)$ itself is derived from $\Gamma(k) \widetilde {H}^{t+\tau}_A(k)$, it follows that $\rho_3 = \Gamma(k)$.

\section{Simulation Results} 

\begin{figure*}
	\centering
	\subfigure[Comparisons of the correlation factors $\rho_1$, $\rho_2$, and $\rho_3$.]{
		\includegraphics[height=4.7cm, width=5.7cm]{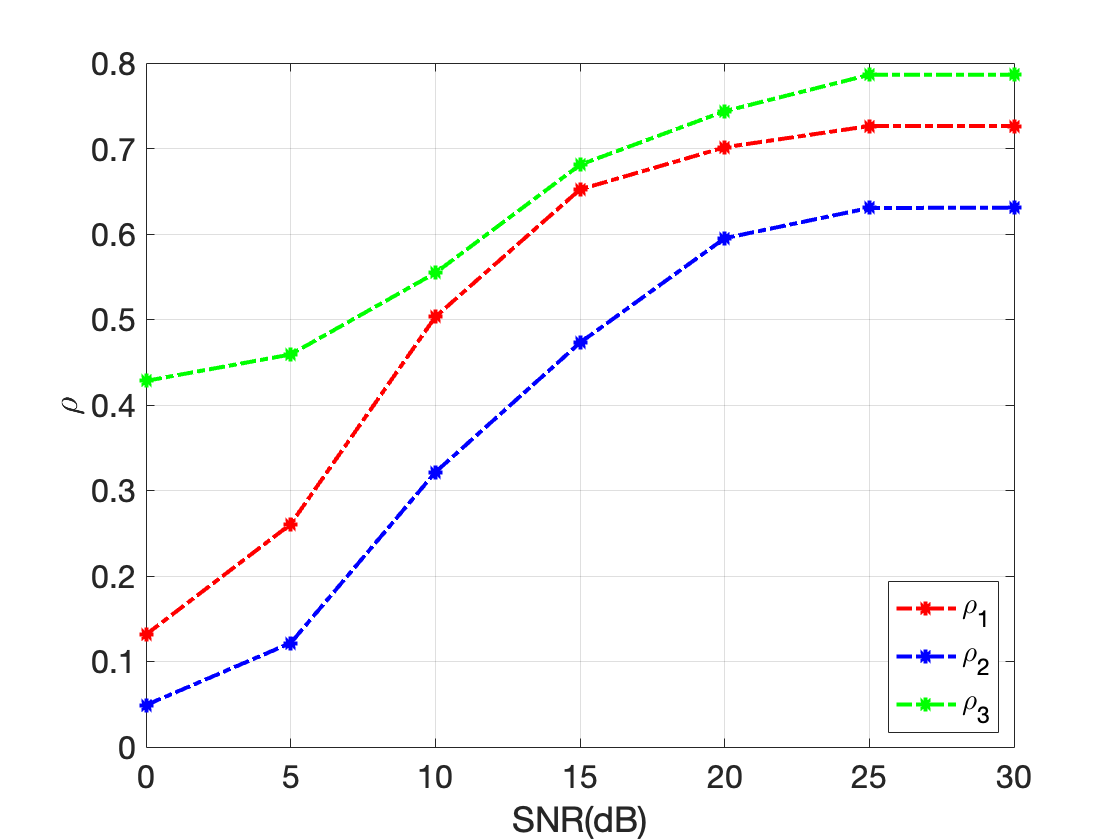}
                \label{5-1}
	}
        \subfigure[MSE predictions  $E\varepsilon(k)^2$ under different RIS matrix uncorrelation with our LoCKey.]{
		\includegraphics[height=4.7cm, width=5.7cm]{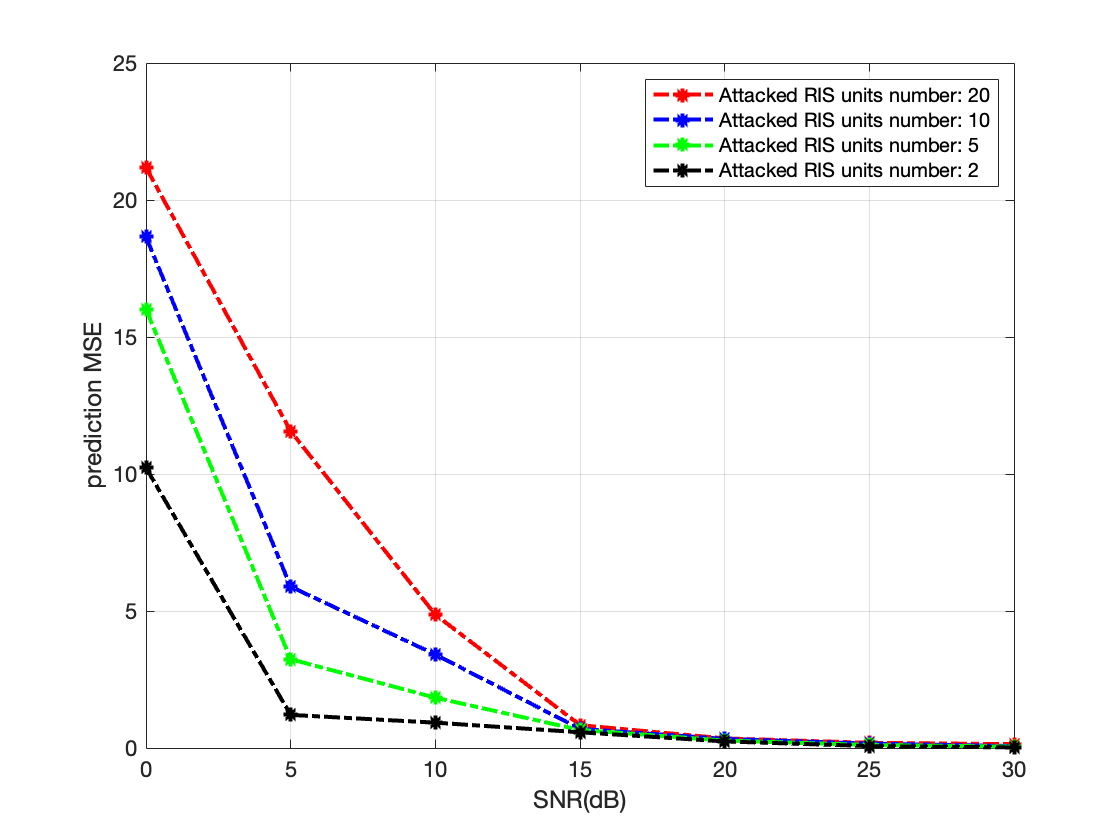}
                \label{5-2}
	}
 \subfigure[$C_{SK}$ comparisons under non loop-back, traditional loop-back and our LoCKey.]{
		\includegraphics[height=4.7cm, width=5.7cm]{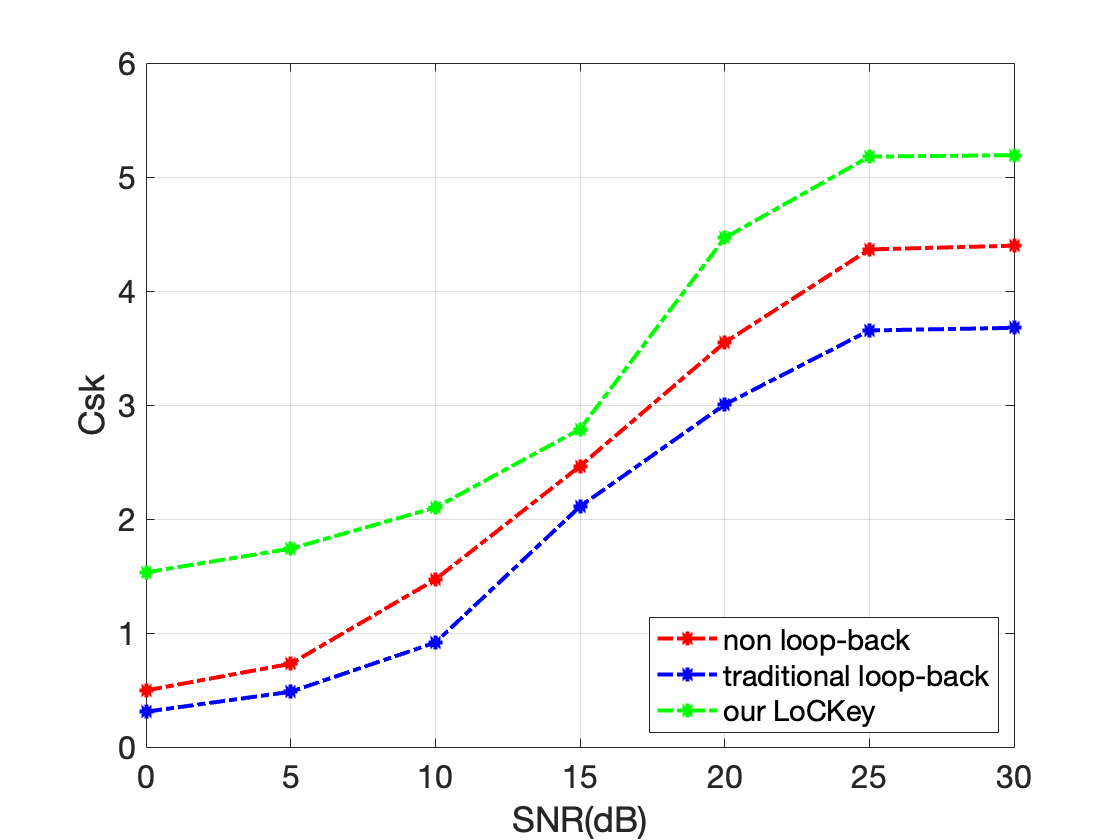}
                \label{5-3}
	}
	\quad
	\subfigure[$C_{SK}$ comparisons under different RIS unit numbers with our LoCKey.]{
		\includegraphics[height=4.7cm, width=5.7cm]{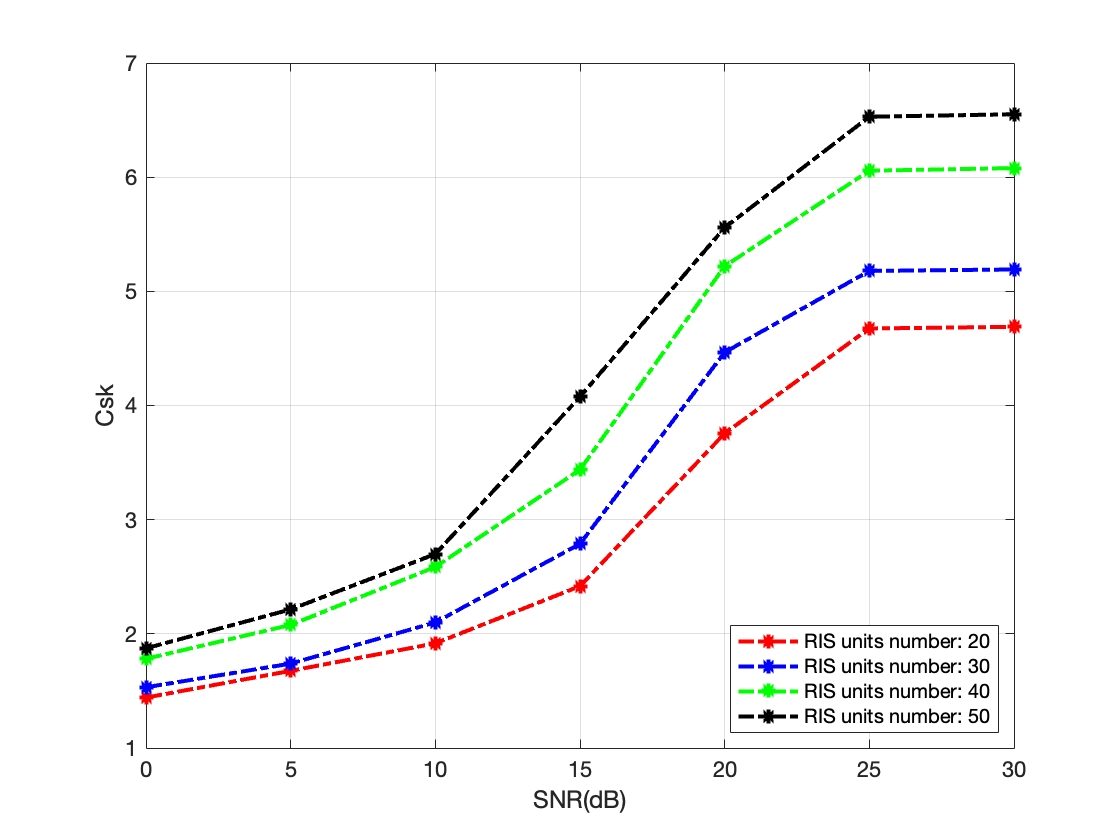}
                \label{5-4}
	}
	\subfigure[$C_{SK}$ comparisons under different RIS matrix uncorrelation with our LoCKey.]{
		\includegraphics[height=4.7cm, width=5.7cm]{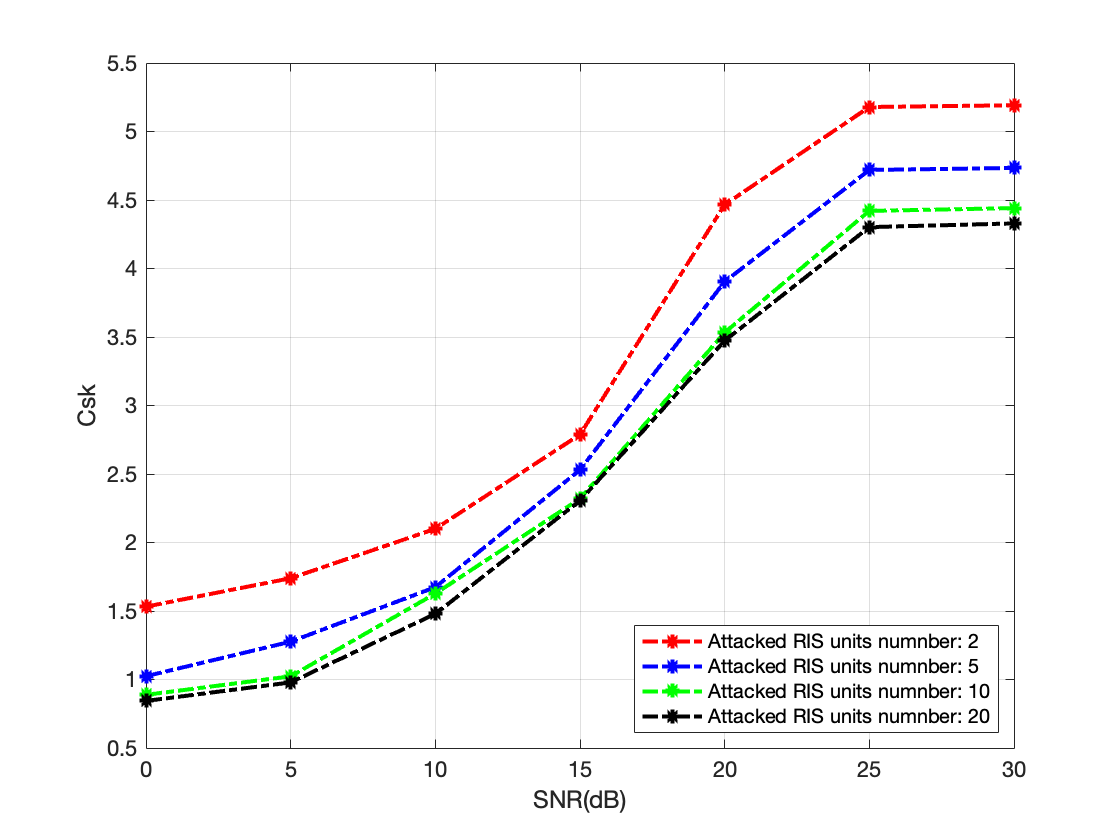}
                \label{5-5}
	}
	\subfigure[Key disagreement rate comparisons under non loop-back, traditional loop-back and our LoCKey.]{
		\includegraphics[height=4.6cm, width=5.3cm]{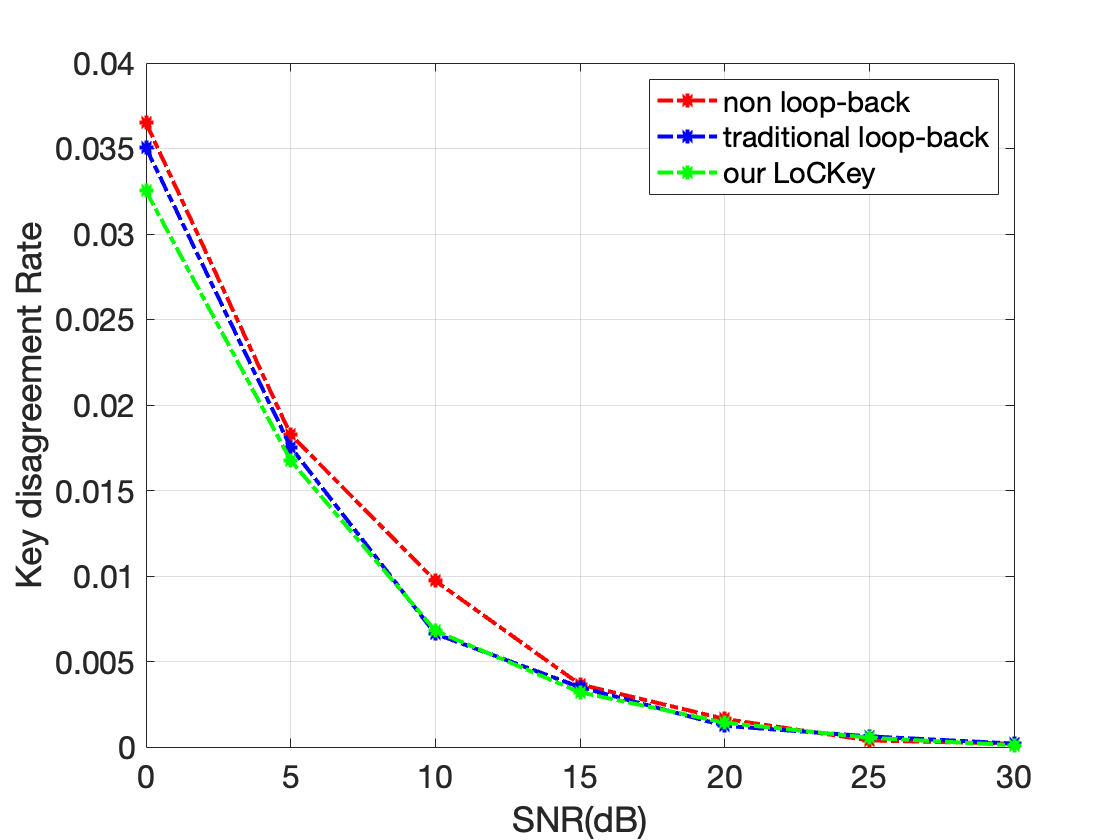}
                \label{5-6}
	}
	\centering
	\caption{Simulation results on our LoCKey scheme. (a) shows the comparisons of the correlation factor before loop-back $\rho_1$, after loop-back $\rho_2$, and after our LoCKey scheme $\rho_3$. (b) shows the MSE predictor value $E[\varepsilon(k)]^2$ after we apply the MMSE compensation module. (c) and (d) show the key generation rate influence brought by different RIS matrix uncorrelation situations and different RIS units numbers, (e) and (f) are the key generation rate and key disagreement rate expression comparisons under non loop-back, traditional loop-back and our LoCKey scheme with the same RIS-jamming attack scenario.}
	\label{5}
\end{figure*}

In order to verify the performance of the proposed LoCKey scheme and compare it with the classical loop-back scheme and non loop-back scheme, extensive simulation experiments are carried out in this section. 

Tapped Delay Line (TDL) channel model is introduced to simulate the communication between Alice-RIS, RIS-Bob and Alice-Bob, hardware fingerprint interference(HF) is modeled as filter with stable multiple taps. 

Table 1 lists the main parameters related to path delay and the average power of each path. Table 2 shows the parameters of detection signal generation in OFDM system.  

\begin{table}[htbp]
\caption{Channel and HF simulation parameters}
\begin{center}

\begin{tabular}{|c|c|c|}
\hline
\textbf{Channel} & \textbf{multipath delay(ms)} & \textbf{average power per path(dB)}\\
\hline
Alice-RIS & [0 0.22 0.50 1.09 1.78] & [0 -4.0 -5.2 -7.0 -1.9]\\
\hline
RIS-Bob& [0 0.37 0.50 1.73 2.82]& [0 -3.0 -5.2 -8.0 -12.2]\\
\hline
Alice-Bob& [0 0.11 0.57 1.90 2.51]& [0 -2.2 -10.5 -6.6 –10.8]\\
\hline
Alice HF & [0 0.13 0.185] & [0 -4 -10]\\
\hline
Bob HF & [0 0.065 0.185] & [0 -7 -10]\\
\hline
\end{tabular}
\label{tab1}
\end{center}
\end{table}

\begin{table}[htbp]
\caption{OFDM basic simulation parameters}
\begin{center}
\begin{tabular}{|c|c|}
\hline
\textbf{Parameters} & \textbf{Value} \\
\hline
OFDM symbol length & 64 \\
\hline
Subcarrier frequency spacing& 15 KHz\\
\hline
Bandwidth& 20 MHz\\
\hline
Carrier frequency at BAND1& 1.82 GHz\\
\hline
Carrier frequency at BAND2& 1.84 GHz\\
\hline
modulation mode & QPSK\\
\hline
cyclic prefix length & 16\\
\hline
pilot interval & 5\\
\hline
iteration number & 1000\\
\hline
maximum doppler shift & 5Hz\\
\hline
RIS units number & 30\\
\hline
\end{tabular}
\label{tab2}
\end{center}
\end{table}

\subsection{The expression of correlation factor $\rho$}

Fig. \ref{5-1} shows that our approach significantly improves the correlation as a downlink CSI value. Interestingly, the traditional loop-back approach is even less relevant than the approach without a loop-back, because although the hardware interference difference is compensated, the loop-back introduces an excess RIS uncertainty that is subject to interference attacks. Our MMSE compensation module can successfully offset this uncertainty, thus achieving a significant improvement in CSI correlation over the two baselines, providing a better key source for subsequent key generation.

\subsection{The expression of prediction MSE $E[\varepsilon(k)]^2$}

As can be seen from Fig. \ref{5-2}, the stronger the interference attack on RIS, the greater the MSE prediction value. For example, when 20 RIS units of the total 30 RIS units are attacked, the MSE prediction value required by our LoCKey under any SNR is significantly higher than that when other RIS units are less likely to be attacked. This means that the worse the reciprocity, the stronger the compensation effect of LoCKey is needed.

\subsection{The improvement brought to key generation rate}

The key generation rate (same meaning with secret key capacity, $C_{SK}$) is defined as the total generated key bit divided by the number of total used subcarriers. We conduct three experiments to evaluate the system performance on $C_{SK}$. 

First, we compare the $C_{SK}$ of LoCKey with the traditional loop-back and non loop-back method when RIS units number is 30 and interfered RIS units number is 5 in Fig. \ref{5-3}. The simulation results show that the $C_{SK}$ of traditional loop-back is lower than that of non loop-back, which is consistent with the channel correlation curve. Our LoCKey is significantly superior to the two baselines, especially when the SNR is low (SNR$<$10dB).

We also compare the effects of our approach on $C_{SK}$ with different RIS element counts and different levels of interference against RIS. It can be inferred from Fig. \ref{5-4} that LoCKey plays a positive role under different numbers of RIS units, the more RIS units there are, the greater the $C_{SK}$ growth rate will be. When the SNR reaches a higher level, the key generation speed tends to be stable, because the performance provided by RIS reaches the previous level. 

From Fig. \ref{5-5}, we learn that our LoCKey can counter interference attacks against different RIS uncorrelation situations. By varying different interference attack levels against the RIS, our LoCKey can still ensure that the KGR reaches a relatively high level under the same SNR conditions, even if the interference rate is 2/3, $C_{SK}$ still reaches 70\% of the interference rate of 1/15.

\subsection{The improvement brought to key consistency}

In addition to a significant increase of $C_{SK}$, our proposed channel compensation method can also significantly improve key consistency. Key disagreement rate (KDR) is defined as the total number of inconsistent key bits between two users divided by the total number of generated key bits. We use gray code for 2-bit quantization. Since there is only 1 bit of difference between adjacent code words in gray code, the inconsistency of the quantization sequence can be effectively limited. As can be seen from Figure \ref{5-6}, our LoCKey has a lower key disagreement rate under any SNR condition, especially when the SNR is lower than that of the traditional loop-back method.

\section{Conclusion}

This paper proposes LoCKey, a novel approach that relies on a loop-back compensation scheme to improve channel reciprocity for physical layer key generation, aiming to effectively mitigate hardware interference and RIS-jamming attacks over open wireless channels. More specifically, the proposed LoCKey involves a novel MMSE compensation module with a signal loop-back strategy to strengthen the overall CSI reciprocity between endpoints. Experimental results show the superiority of our method.

\section{Acknowledgments}
This work was supported by the National Key R\&D Program of China (No. 2022YFB2902200).

\end{document}